\documentclass[amsmath,amssymb,aps,letterpaper,prl,twocolumn,longbibliography]{revtex4-1}

\usepackage{graphicx}
\usepackage{color}
\usepackage{physics}
\usepackage{hyperref}
\usepackage{mathtools}

\usepackage{enumitem}

\usepackage{picture}
\usepackage{calc}

\usepackage{subfigure}
\usepackage{dsfont}
\usepackage{color}

\usepackage[utf8x]{inputenc}

\usepackage{graphicx}
\usepackage{tikz}

\usetikzlibrary{decorations.markings}

\tikzset{line/.style={line width=0.25mm},
curve/.style={line,smooth,tension=1},
->-/.style={decoration={
  markings,
  mark=at position #1 with {\arrow[>=stealth]{>}}},postaction={decorate}},
-<-/.style={decoration={
  markings,
  mark=at position #1 with {\arrow[>=stealth]{<}}},postaction={decorate}},
}

\begin{document}

\title{Integrable and critical Haagerup spin chains}

\author{Luke Corcoran}%
 \email{lcorcoran@maths.tcd.ie}
\author{Marius de Leeuw}%
 \email{mdeleeuw@maths.tcd.ie}

\affiliation{%
School of Mathematics
\& Hamilton Mathematics Institute\\
Trinity College Dublin, Ireland
}%

\begin{abstract}
We construct the first integrable models based on the Haagerup fusion category $H_3$. We introduce a Haagerup version of the anyonic spin chain and use the boost operator formalism to identify two integrable Hamiltonians of PXP type on this chain. The first of these is an analogue of the golden chain; it has a topological symmetry based on $H_3$ and satisfies the Temperley-Lieb algebra with parameter $\delta=(3+\sqrt{13})/2$. We prove its integrability using a Lax formalism, and construct the corresponding solution to the Yang--Baxter equation. We present numerical evidence that this model is gapless with a dynamical critical exponent $z\neq 1$. The second integrable model we find breaks the topological symmetry. We present numerical evidence that this model reduces to a CFT in the large volume limit with central charge $c\sim3/2$.
\end{abstract}

\maketitle

\section{Introduction}

Non-invertible symmetries have played a prominent role in recent years, with applications ranging widely from condensed matter systems and critical phenomena to topological lines in quantum field theory \cite{McGreevy:2022oyu,Cordova:2022ruw,Bhardwaj:2023kri,Schafer-Nameki:2023jdn,Shao:2023gho}. They appear notably in the context of the 2d Ising model via the Kramers-Wannier duality \cite{PhysRev.60.263}, a generalisation of which is given by the statistical AFM model \cite{Aasen:2016dop, Aasen:2020jwb}. By taking the anisotropic limit of this model, one can obtain one-dimensional quantum anyonic chains \cite{Inamura:2023qzl}. These models also enjoy non-invertible symmetries and are based on the data of fusion categories \cite{fusion}. 

Fusion categories are non-invertible extensions of groups, and consist of a set of simple objects $a,b,c,\dots$, which can also be thought of as anyon species. These simple objects can be combined using the fusion rule
\begin{equation}
a\times b= \bigoplus_{c} N_{ab}^c \hspace{0.1cm}c,
\end{equation}
where $N_{ab}^c$ are non-negative integers. The associativity of the usual group law is relaxed using the data of $F$-symbols:
\begin{equation}
\begin{gathered}
\begin{tikzpicture}[scale=.3]
\draw [line] (-4,2) node [above] {$a$} -- (-3,1) node [below,xshift=0.4mm,yshift=-0.5mm] {$p$} -- (-2,0) -- (-2,-1) node [below] {$u$};
\draw [line] (-3,1) -- (-2,2) node [above] {$b$};
\draw [line] (-2,0) -- (0,2) node [above] {$c$};
\node at (1,0){$=$};
\node at (5,0){$\sum_{q}(F_{u}^{abc})_{qp}$};

\draw [line] (8,2) node [above] {$a$}-- (10,0) -- (10,-1) node [below] {$u$};
\draw [line] (10,0) -- (12,2) node[above]{$c$};
\draw [line] (11,1) node [below,xshift=-0.4mm,yshift=-0.4mm] {$q$} -- (10,2) node [above]{$b$};

\end{tikzpicture}
\end{gathered},
\end{equation}
where $(F_{u}^{abc})_{qp}\in \mathbb{C}$. These $F$-symbols need to satisfy a set of consistency conditions which arise from 4 anyon fusion known as the pentagon equations. They are the key input for the construction of fusion category symmetric anyonic chains.

The simplest example is given by that of the golden chain \cite{Feiguin:2006ydp}, which is related to the Fibonacci fusion category \cite{Trebst_2008}. This model has many interesting properties; it is a critical lattice model which corresponds to CFTs at both ends of its spectrum in the continuum limit. It is also an integrable model, since it is a special point of an integrable spin chain related to Baxter's hard square model \cite{Fendley_2004, Baxter-Book}. This integrable model is related to the RSOS models of Andrews, Baxter, and Forrester \cite{RSOS-1,RSOS-2}, see also \cite{RSOS-H}. The spin-chain Hilbert space for the golden chain is \textit{constrained}; in particular it is a spin-1/2 chain where neighbouring down spins are disallowed. This constraint is known as the `Rydberg blockade' \cite{Urban_2009}. The Rydberg-constrained Hilbert space admits several interesting models which do not commute with the fusion category symmetry, for example the PXP model \cite{PhysRevA.86.041601,Serbyn:2020wys,PhysRevB.98.155134}, the Lesanovsky model \cite{lesanovsky,lesanovsky-model,pxp-int}, and the constrained XXZ model \cite{constrained1,constrained2}.

It is a natural question whether there are interesting anyonic chains based on fusion categories beyond the Fibonacci case. The simplest example of a fusion category which is not simply related to (affine) Lie algebras or quantum groups is the so-called Haagerup fusion category \cite{haagerup,Grossman_2012}. The $F$-symbols were calculated recently \cite{osborne2019fsymbols,Huang:2020lox}, which allowed for the investigation of Haagerup symmetric models at the level of an anyonic chain \cite{Wolf:2020qdo,Huang:2021nvb} and a 2d lattice model \cite{Vanhove:2021zop}. Both \cite{Huang:2021nvb,Vanhove:2021zop} identify critical models with $c=2$, although the relation between these models is still unclear.

In this letter we study Haagerup anyonic spin chains from the point of view of integrability, which provides a new way to generate potentially interesting models. In recent years, the boost operator formalism has proven an effective method for generating integrable models on various Hilbert spaces \cite{marius-classification-1,marius-classification-2,marius-classification-3,marius-classification-4,marius-classification-5,marius-classification-6}. This method has recently been adapted to higher range models \cite{Gombor:2021nhn} and constrained Hilbert spaces \cite{Corcoran:2024ofo}. We apply this framework to partially classify the set of range 3 integrable models on the Haagerup-constrained Hilbert space. We identify one of the models obtained with the projector onto the identity fusion channel, $\mathsf{P}_1$, so that this model can be seen as one generalisation of the golden chain to the Haagerup Hilbert space. The alternative generalisation, $\mathsf{P}_{\rho}$, is critical but notably not integrable. The second model we find partially breaks the Haagerup topological symmetry, but appears to be a critical spin chain in the large volume limit with central charge $c\sim 3/2$.

\section{Haagerup Hilbert space}
The Haagerup fusion category $H_3$ consists of six simple objects $\mathrm{Obj}(H_3)=\{1,a,a^2,\rho, a\rho, a^2\rho\}$. The non-trivial fusion rules are
\begin{equation}
 \rho \times a^2 =a\rho, \qquad \rho\times \rho = 1 + \rho + a\rho + a^2\rho,
\end{equation}
and $a^3 = 1$. An anyonic chain based on this fusion category can be constructed, which describes the interaction of $L$ $\rho$-anyons. We denote the Hilbert space by $V^L$, and the basis states are given via the fusion diagram \cite{Feiguin:2006ydp,Huang:2021nvb}
\begin{equation}
    \label{eq:Basis}
\ket{x_1x_2\cdots x_L}=
\begin{gathered}
\begin{tikzpicture}[scale=.8]
\draw [line] (-1,0) node {$/\!/~$} -- (-.5,0) node [below] {$x_1$} -- (.5,0) node [below] {$x_2$}
-- (1.5,0) node [below] {$\dotsb$}
-- (2.5,0) node [below] {$x_L$} -- (3.5,0) node [below] {$x_1$} -- (4,0) node {$~/\!/$};
\draw [line] (0,1) node [above] {$\rho$} -- (0,0);
\draw [line] (1,1) node [above] {$\rho$} -- (1,0);
\draw [line] (2,1) node [above] {$\rho$} -- (2,0);
\draw [line] (3,1) node [above] {$\rho$} -- (3,0);
\end{tikzpicture}
\end{gathered},
\end{equation}
where $x_i\in\mathrm{Obj}(H_3)$ and slashed lines represent periodic identification. In order for \eqref{eq:Basis} to represent an allowed state in the Hilbert space, it must correspond to a valid fusion diagram in $H_3$. In other words, the fusion $x_i\times \rho$ must contain $x_{i+1}$, for each $i=1,2,\dots,L$.

\begin{figure}
\begin{tikzpicture}[scale=0.9]

\draw (-1.8,0) circle [radius=0.4] node {$a \rho$} ;
\draw (-0.7,0) circle [radius=0.4] node {$a $};
\draw (0.7,0) circle [radius=0.4] node {$a^2 $};
\draw (1.8,0) circle [radius=0.4] node {$a^2\rho$};
\draw (0,1) circle [radius=0.4] node {$1$};
\draw (0,2.1) circle [radius=0.4] node {$\rho$};

\draw [line,] (-1.1,0) -- (-1.4,0);
\draw [line,] (1.1,0) -- (1.4,0);
\draw [line,] (0,1.4) -- (0,1.7);

\draw [line] (-1.535, 0.309167) -- (-0.265, 1.79083);
\draw [line] (1.535, 0.309167) -- (0.265, 1.79083);

\draw[line]    (-1.5172,-0.2828) to[out=-30,in=-150] (1.5172,-0.2828);

\draw[line]    (-1.8-0.2828,0.2828) to[out=-210,in=-150,looseness=7] (-1.8-0.2828,-0.2828);

\draw[line]    (1.8+0.2828,0.2828) to[out=30,in=-30,looseness=7] (1.8+0.2828,-0.2828);

\draw[line]    (0.2828,2.1+0.2828) to[out=60,in=120,looseness=7] (-0.2828,2.1+0.2828);

\end{tikzpicture}
\caption{Adjacency rules for the Haagerup Hilbert space. Nodes represent simple objects in $H_3$. If there is no edge between a pair of objects then this is a disallowed pair in $V^L$.}
\label{fig:allowedstate}
\end{figure}
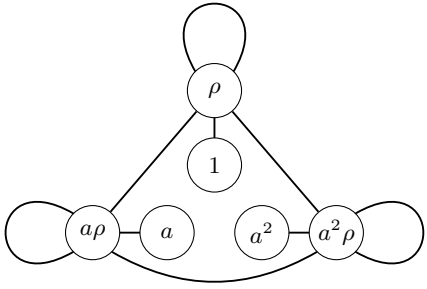

$V^L$ can be obtained from $(\mathbb{C}^6)^{\otimes L}$ by projecting away disallowed neighbouring pairs. Of the $36$ potential neighbouring pairs in $\mathbb{C}^6\otimes \mathbb{C}^6$, only 15 are allowed by the fusion diagram \eqref{eq:Basis}. The precise adjacency rules are summarised in figure \ref{fig:allowedstate}.
We introduce an explicit projector 
\begin{equation}
\Pi^L = \prod_{i=1}^L \Pi_{i,i+1}, \qquad \Pi^L: (\mathbb{C}^6)^{\otimes L}\rightarrow V^L,
\end{equation}
where $\Pi_{i,i+1}\in \text{End}(\mathbb{C}^6\otimes \mathbb{C}^6)$ projects away disallowed neighbouring states. Due to this constraint the Hilbert space $V^L$ does not have a natural tensor product form. Its dimension grows as $d_L \sim \psi^L$, where $\psi=(3+\sqrt{13})/2$.

\section{Non-invertible Symmetry}
An operator $\mathcal{O}:V^L\rightarrow V^L$ has Haagerup symmetry if it commutes with a set of non-local operators $Y_z$, $z\in \mathrm{Obj}(H_3)$. The action of $Y_z$ is defined by fusing $z$ into a state $\ket{x_1x_2\cdots x_L}$ below the chain \cite{Feiguin:2006ydp,Gils_2013,Buican:2017rxc}. Its matrix elements are given explicitly by
\begin{equation}
\bra{y_1y_2\cdots y_L}Y_z \ket{x_1x_2\cdots x_L} =  \prod_{i=1}^L (F_{y_{i+1}}^{zx_i\rho})_{x_{i+1}y_i},
\end{equation}
where $(F_{u}^{abc})_{qp}$ are the $F$-symbols of $H_3$ \cite{osborne2019fsymbols}. The operators $Y_z$ furnish a representation of the fusion algebra, and therefore $\mathcal{O}$ only needs to commute with $Y_a$ and $Y_\rho$ to have Haagerup symmetry. 

We choose to use an $F$-symbol gauge where $Y_a$ coincides with the operator fusing all $x\in \mathrm{Obj}(H_3)$ with $a$. In other words, $Y_a$ simply rotates the labels $1\rightarrow a\rightarrow a^2\rightarrow 1$ and $\rho\rightarrow a\rho \rightarrow a^2\rho \rightarrow \rho$ \footnote{In practise we achieve this by using the $F$-symbols of \cite{osborne2019fsymbols} with a mild gauge transformation.}.

We note that there are a set of range 3 operators $\mathsf{P}_z$ for $z\in \{1,\rho,a\rho,a^2\rho\}$, which trivialise the Haagerup symmetry on $V^L$. $\mathsf{P}_z$ is given by the projection of a pair of $\rho$ objects onto the object $z$ \cite{Wolf:2020qdo,Huang:2021nvb}. In terms of $F$-symbols, these projectors take the form $\mathsf{P}_z=-\sum_{i=1}^L \mathsf{P}_z^{(i)}$, where
\begin{align}\label{eq:HagProj}
&\mathsf{P}_{z}^{(i)}: \ket{x_{i-1}x_ix_{i+1}}=  \\ \notag &\sum_{x_i'\in \mathrm{Obj}(H_3)}(F_{x_{i+1}}^{x_{i-1}\rho\rho})_{zx_i}(F_{x_{i+1}}^{x_{i-1}\rho\rho})_{zx_i'}\ket{x_{i-1}x_i'x_{i+1}}.
\end{align}
The critical model studied in \cite{Huang:2021nvb} is $\mathsf{P}_\rho$. We note that these projectors are not all independent; they satisfy the relation
$\mathsf{P}_1 + \mathsf{P}_\rho+\mathsf{P}_{a\rho}+\mathsf{P}_{a^2 \rho} = 1$.

\section{Integrability}
Here we are searching for \textit{integrable} Hamiltonians $\mathbb{Q}_2^{(\Pi)}: V^L\rightarrow V^L$ on the Haagerup Hilbert space, using the constrained boost operator formalism of \cite{Corcoran:2024ofo}. We call a range 3 Hamiltonian $\mathbb{Q}_2:(\mathbb{C}^6)^{\otimes L}\rightarrow (\mathbb{C}^6)^{\otimes L}$ integrable on $V^L$ if it is a member of an infinite tower of constraint compatible operators $\mathbb{Q}_i:(\mathbb{C}^6)^{\otimes L}\rightarrow (\mathbb{C}^6)^{\otimes L}$ which mutually commute on the constrained subspace:
\begin{equation}\label{eq:QiQj}
\Pi^L[\mathbb{Q}_i, \mathbb{Q}_j] \Pi^L = 0, \qquad [\mathbb{Q}_i, \Pi^L]=0.
\end{equation}
For translationally invariant spin chains the first charge $\mathbb{Q}_1$ can always be taken as the momentum operator, which generates translations along the chain. For Yang--Baxter integrable spin chains, $\mathbb{Q}_3$ can be obtained from the Hamiltonian using a higher-range analogue of the so-called \textit{boost operator} \cite{1982JETP...55..306T,links2001ladder,Gombor:2021nhn}:
\begin{equation}\label{eq:Q3}
\mathbb{Q}_3=\sum_{i=1}^{L} [\mathcal{H}_{i,i+1,i+2},\mathcal{H}_{i+1,i+2,i+3}+\mathcal{H}_{i+2,i+3,i+4}],
\end{equation}
where $\mathbb{Q}_2=-\sum_{i=1}^L \mathcal{H}_{i,i+1,i+2}.$
Therefore, given a Hamiltonian density $\mathcal{H}$, a necessary condition for integrability is
\begin{equation}\label{eq:Q2Q3}
\Pi^L[\mathbb{Q}_2, \mathbb{Q}_3] \Pi^L = 0, 
\end{equation}
where $\mathbb{Q}_3$ is constructed from $\mathbb{Q}_2$ using \eqref{eq:Q3}. This condition is believed to be sufficient; we are not aware of any models satisfying \eqref{eq:Q2Q3} which are not integrable. To prove integrability it is necessary to construct a Lax operator which generates the charges $\mathbb{Q}_i$ and a corresponding $R$-matrix.

\section{Hamiltonian ansatz}
We search for solutions to the integrability condition \eqref{eq:Q2Q3}, and so require an appropriate ansatz for the Hamiltonian density $\mathcal{H}$. We require that $\mathbb{Q}_2$ is symmetric and invariant under spatial reflection. We further require that $\mathcal{H}_{i,i+1,i+2}$ takes a generalised `PXP' form
\begin{equation}
\mathcal{H}_{i,i+1,i+2} = D_i \mathcal{O}_{i+1} D_{i+2},
\end{equation}
where $D:\mathbb{C}^6\rightarrow \mathbb{C}^6$ is diagonal. Finally, we require partial Haagerup symmetry, and impose that $\mathbb{Q}_2^{(\Pi)}\coloneqq \Pi^L \mathbb{Q}_2$ commutes with $Y_a$. We trivialise $Y_a$ symmetry by decomposing the Hamiltonian density
\begin{equation}
\mathcal{H}_{i,i+1,i+2}=\tilde{ \mathcal{H}}_{i,i+1,i+2}+\tilde{\mathcal{H}}^a_{i,i+1,i+2}+\tilde{\mathcal{H}}^{a^2}_{i,i+1,i+2},
\end{equation}
where $\tilde{\mathcal{H}}^a_{i,i+1,i+2}=Y_a \tilde{ \mathcal{H}}_{i,i+1,i+2} Y_a^{-1}$. After imposing these requirements, there are 18 operators contributing to the the reduced Hamiltonian $\tilde{\mathcal{H}}_{i,i+1,i+2}$. These operators can be written in terms of the following local operators.

\paragraph{Diagonal operators.} The diagonal operators are spanned by the projectors $P_x$, for $x\in\mathrm{Obj}(H_3)$, which act via $P_x \ket{y} = \delta_{xy}\ket{y}.$

\paragraph{Flips.} We will denote by $X,Y,Z$ the operators which flip between $\ket{1}$ and $\ket{\rho}$, $\ket{a\rho}$, and $\ket{a^2\rho}$ respectively:
\begin{equation}
X: \ket{1}\leftrightarrow \ket{\rho}, \hspace{0.32cm} Y: \ket{1}\leftrightarrow \ket{a\rho},  \hspace{0.32cm}Z: \ket{1}\leftrightarrow \ket{a^2\rho}.
\end{equation}

\paragraph{Transpositions.} There is one more independent operator which transposes non-invertible objects:
\begin{equation}
T: \ket{\rho}\leftrightarrow \ket{a\rho}.
\end{equation}
\section{Integrable Solutions}

To search for integrable Hamiltonians on $V^L$, we use a Hamiltonian ansatz which is a linear combination of the 18 operators contributing to $\tilde{\mathcal{H}}_{i,i+1,i+2}$. We use this to form the higher charge $\mathbb{Q}_3$ using \eqref{eq:Q3}, and impose the integrability condition \eqref{eq:Q2Q3}. This leads to a set of highly coupled cubic equations in the coefficients of $\mathcal{\tilde{H}}$. 
\paragraph{Model 1.} We are able to fully classify the solutions of these equations which further commute with the symmetry operator $Y_\rho$. We find a single solution, where the reduced Hamiltonian takes the form
\begin{align}\label{eq:Hsol}
&\tilde{\mathcal{H}}_{i,i+1,i+2}= \psi^{-1/2} P_{\rho}(X+Y+Z)P_\rho + P_{\rho}TP_{\rho} \notag \\
+&P_{a\rho}TP_{a\rho}+ P_{a^2\rho}TP_{a^2\rho} + \psi  P_1P_{\rho}P_1 + \psi^{-1}P_\rho P_1 P_\rho \notag \\
+& P_\rho P_\rho P_\rho+ P_{\rho}P_{a\rho}P_{\rho}+P_{a\rho}P_{\rho}P_{a\rho},
\end{align}
where for brevity we omitted indices on the right hand side. For example, the first term should be $ \psi^{-1/2} P_{\rho,i} X_{i+1} P_{\rho,i+2}$. This Hamiltonian coincides with $\mathsf{P}_1$, as given by \eqref{eq:HagProj}.

\paragraph{Model 2.}
We partially classified the solutions of \eqref{eq:Q2Q3} which do not commute with $Y_\rho$. One solution is
\begin{align}\label{eq:Hsol2}
\tilde{\mathcal{H}}_{i,i+1,i+2}&= \gamma^{-1/2} P_{\rho}(X+Y+Z)P_\rho\notag \\ +& P_{\rho}TP_{\rho} +P_{a\rho}TP_{a\rho}+ P_{a^2\rho}TP_{a^2\rho}\notag\\
 +&(1+2\gamma) P_1P_{\rho}P_1 + (1+2\gamma^{-1})P_\rho P_1 P_\rho \notag \\
+&2\gamma( P_\rho P_\rho P_\rho+ P_{\rho}P_{a\rho}P_{\rho}+P_{a\rho}P_{\rho}P_{a\rho}),
\end{align}
where $\gamma=(1+\sqrt{3})/2$.

\section{Lax Operator and Yang--Baxter}
The Hamiltonians \eqref{eq:Hsol} and \eqref{eq:Hsol2} obey the necessary condition \eqref{eq:Q2Q3} for integrability. In order to prove integrability, we need to exhibit both as a member of a larger family of commuting charges $\mathbb{Q}_i$. We do this by the construction of a \textit{Lax operator} $\mathcal{L}_{Aj}(u)\in \text{End}(V_A \otimes V_j)$. $V_A\simeq \mathbb{C}^6\otimes \mathbb{C}^6$ is the \textit{auxiliary space} \footnote{For models of range $r$, the auxiliary space contains $r-1$ copies of the physical space.}, $V_j\simeq \mathbb{C}^6$ is a copy of the local physical space, and $u$ is a spectral parameter. The Lax operator is used to form a transfer matrix, which is a generating function for the set of commuting charges.
\paragraph{Model 1.}
Writing $\mathbb{Q}_2 = -\sum_{i=1}^L e_i$ we find that the generators $e_i$ satisfy the Temperley-Lieb algebra
\begin{align}\label{eq:TL}
&e_i^2 = \psi \hspace{0.05cm} e_i, \qquad e_ie_{i\pm 1}e_i = e_i, \notag \\
&[e_i,e_j]=0, \qquad |i-j|>1.
\end{align}
In such cases the Lax operator can typically be obtained as a linear combination of the identity matrix and $\mathcal{H}$. Indeed, we find that
\begin{align}\label{eq:Lax2}
\mathcal{L}_{123}(u)&=\mathcal{P}_{13}\mathcal{P}_{23}\left(1-\frac{1}{\frac{\psi}{2}-\alpha \coth{\alpha \hspace{0.03cm}u}} \mathcal{H}_{123}\right)\notag\\ &\coloneqq \mathcal{P}_{13}\mathcal{P}_{23} \check{\mathcal{L}}_{123}(u)
\end{align}
is appropriate, where $\mathcal{P}_{ij}$ is the permutation operator on $V_i\otimes V_j$ and $\alpha\coloneqq \sqrt{\frac{3}{4}(\psi-1)}$. The transfer matrix is defined by
\begin{equation}\label{eq:transfer}
t(u) \coloneqq \tr_A [\mathcal{L}_{AL}(u)\mathcal{L}_{A,L-1}(u)\cdots \mathcal{L}_{A1}(u)],
\end{equation}
and the corresponding charges $\mathbb{Q}_i$ are extracted from $t(u)$ via $i-1$ logarithmic derivatives at $u=0$.

In order to show the integrability condition \eqref{eq:QiQj}, we look for a solution to the constrained R$\mathcal{L}\mathcal{L}$ equation \cite{Corcoran:2024ofo}. We find the following $R$-matrix $R_{AB}(u,v)$:
\begin{equation}\label{eq:Rsol}
\check{R}_{1234}(u,v)=\check{\mathcal{L}}_{123}(-v)\check{\mathcal{L}}_{234}(u-v)\check{\mathcal{L}}_{123}(u),
\end{equation}
 which satisfies appropriate projected RLL and Yang--Baxter equations \cite{Gombor:2021nhn,Corcoran:2024ofo}. This $R$-matrix can be used to prove $\Pi^L[t(u), t(v)]\Pi^L=0$ \footnote{The proof is closely related to the standard `train-track' argument of integrability, see e.g. \cite{faddeev1996algebraic}, which constructs an RTT relation from the RLL relation and proves $[t(u),t(v)]=0$ by taking a trace. This has been adapted in \cite{Corcoran:2024ofo} to integrable models on constrained Hilbert spaces by inserting appropriate projectors.},
from which the integrability condition \eqref{eq:QiQj} follows.
\paragraph{Model 2.}
The Lax operator for model 2 is given by
\begin{equation}
\check{\mathcal{L}}_{123}(u) = 1 + u \mathcal{H}_{123} + f(u) \mathcal{H}_{123}^2,
\end{equation}
where $(u + 2\gamma f(u))^2 = f(u)(f(u)+2)$.
\section{Numerics}
We investigate the low-lying spectrum of the obtained integrable models. For a gapless model the energy gap vanishes at large $L$:
\begin{equation}\label{eq:gap}
\Delta E \coloneqq E_1 - E_0 \propto \frac{1}{L^z}.
\end{equation} 
For a CFT we expect a dynamical critical exponent $z=1$ and the half-chain entanglement entropy of the ground state on the periodic chain to scale as \cite{Holzhey:1994we,Calabrese:2004eu}
\begin{equation}\label{eq:ee}
S \sim \frac{c}{3} \log L.
\end{equation}
We computed both of these quantities numerically for the Hamiltonians \eqref{eq:Hsol} and \eqref{eq:Hsol2}. We studied the periodic chain because integrability is preserved and the ground state is non-degenerate for both models. We used methods of DMRG with the iTensor package \footnote{We used a maximum truncation error of $10^{-9}$ and a maximum bond dimension of 1700.},\cite{Fishman_2022}. We verified these results up to $L=14$ via exact diagonalisation.

\begin{figure}[htbp]
  \begin{minipage}[b]{0.46\textwidth}
    \centering
    \includegraphics[width=\textwidth]{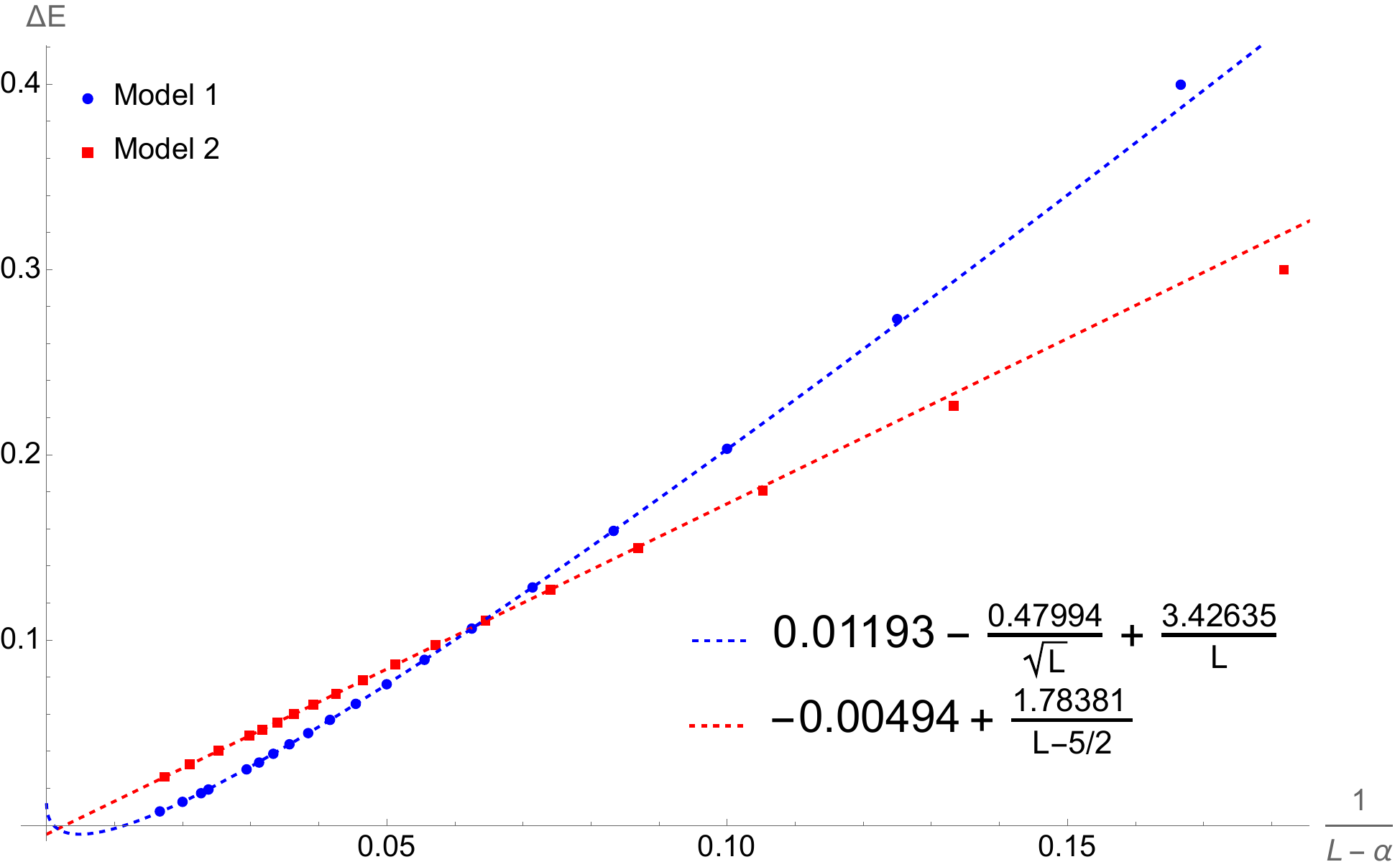}
    \caption{Energy gap up to $L=60$ on the periodic chain vs. an inverse shifted length $(L-\alpha)^{-1}$. For model 1 we take $\alpha=0$ and for model 2 we take $\alpha=5/2$. For the model 2 fit we used the last 10 points.}
    \label{fig:gap}
  \end{minipage}
  \hspace{0.05\textwidth}
   \begin{minipage}[b]{0.46\textwidth}
    \centering
    \includegraphics[width=\textwidth]{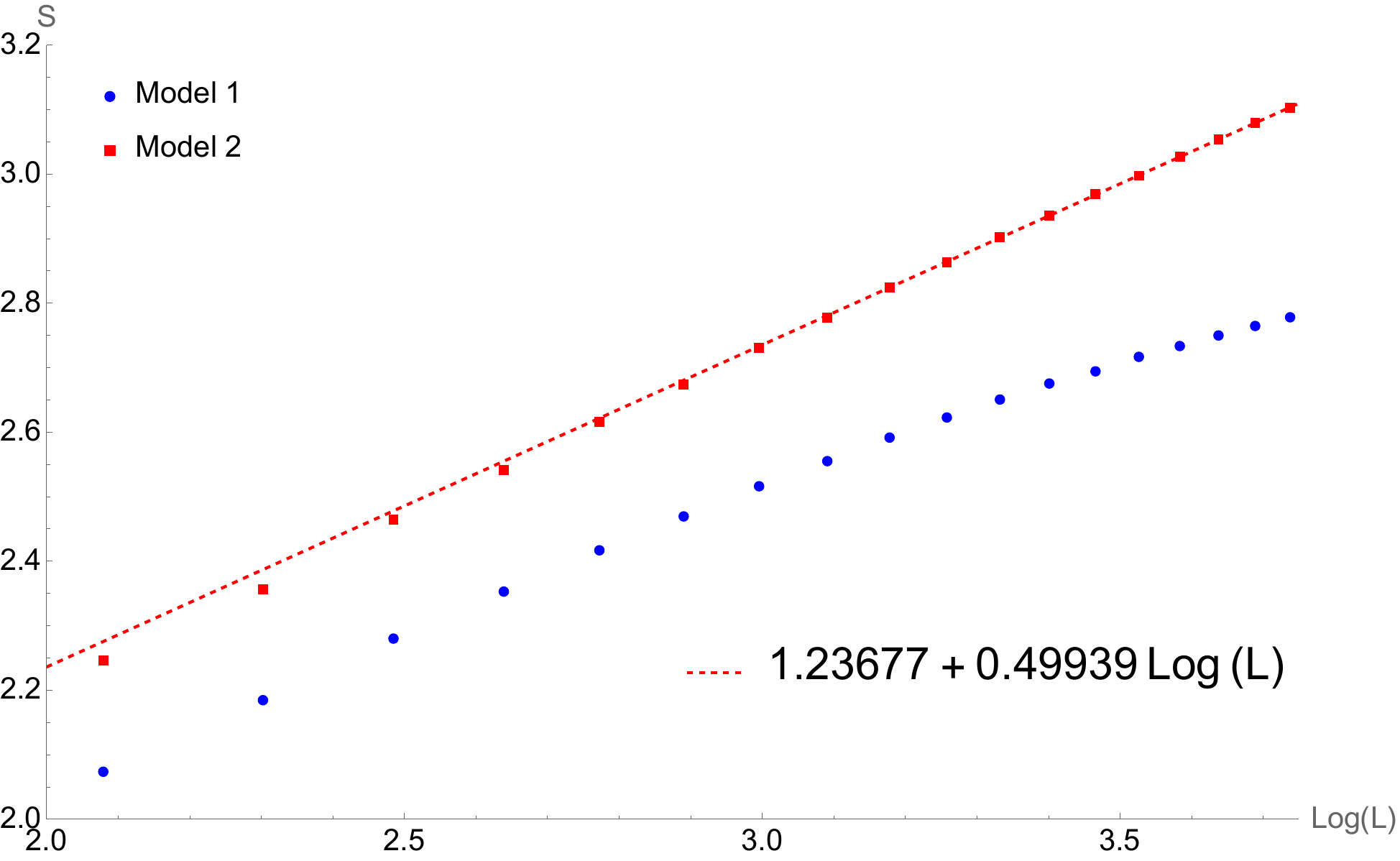}
    \caption{Half-chain entanglement entropy up to $L=42$ on the periodic chain. For the fit we used the last 10 points.}
    \label{fig:ee}
  \end{minipage}
\end{figure}

For model 1 we find evidence of gaplessness. While we were unable to determine the analytic behaviour of the gap, any Laurent series fit in $L$ or $L^{1/2}$ leads to a small or negative gap in the large-$L$ limit. Since the gap needs to be positive, this points towards a vanishing gap. In figure \ref{fig:gap} we present a fit of the gap to $a+bL^{-1/2}+c L^{-1}$, which was one of the best fits we obtained. More data is needed to understand the precise analytic structure. We expect that the model is not a CFT since we find a dynamical critical exponent $z\neq 1$. This is supported by our results of the half-chain entanglement entropy, which does not scale as $\log L$, see figure \ref{fig:ee}.

For model 2 we find evidence that this model is critical in the large volume limit. We find in figure \ref{fig:gap} that the gap scales as $(L-5/2)^{-1}$ as $L\rightarrow \infty$. Furthermore, we find that the half-chain entanglement entropy scaling in figure \ref{fig:ee} is consistent with \eqref{eq:ee} for a central charge $c\sim 3/2$. We see that for $L>20$ both plots become nicely linear. We therefore propose model 2 as the first example of a critical Haagerup spin chain which is integrable.

\section{Conclusions \& Outlook}
In this letter we studied Haagerup spin chains from the perspective of integrability. We applied the boost operator formalism and presented two integrable spin chains on the Haagerup Hilbert space. We find that only one integrable Hamiltonian, $\mathsf{P}_1$, is compatible with the full Haagerup symmetry. The other model we present breaks this topological symmetry, but seems to correspond to a CFT in the continuum limit.

Due to integrability, the Temperley-Lieb algebra \eqref{eq:TL}, and the fact that the model is one generalisation of the golden chain, it was reasonable to expect that \eqref{eq:Hsol} also corresponds to a CFT in the continuum limit. We do not find evidence of this at our current level of numerics. This is potentially surprising, given that many TL models correspond to CFTs. A potential reason in this case is that the Temperley-Lieb parameter $\delta=\psi = q+q^{-1}$ does not correspond to $q$ which is a root of unity, which differs from the golden chain case. Since the model still appears to be gapless, it is interesting to consider whether this TL algebra corresponds at least to some deformation of conformal symmetry.

The golden chain corresponds to the projector $\mathsf{P}_1$ constructed from the Fibonacci fusion category $F$-symbols, or equivalently to $\mathsf{P}_\tau = 1-\mathsf{P}_1$. Therefore both $\mathsf{P}_1$ and $\mathsf{P}_\rho$ can be considered as Haagerup extensions of the golden chain. We see that while $\mathsf{P}_\rho$ inherits criticality from the golden chain, it is $\mathsf{P}_1$ which inherits the integrability properties. We conjecture that this pattern to persists for Haagerup-Izumi fusion categories of higher rank \cite{Huang:2020lox}.

We presented one more integrable model which appears to be critical in the large volume limit, with central charge $c\sim3/2$. Since we find a shifted scaling behaviour $\Delta E\sim 1/(L-\alpha)$, it is important to develop more specific numerical and analytic tools for anyonic spin chains, to push the computation of the gap and other to higher length. This is also necessary to study the ferromagnetic versions of our models, which have highly degenerate ground states.

There are hints that both models we present are different points of a more general family of integrable Hamiltonians on the Haagerup Hilbert space. It would be interesting to identify this parent integrable model, which potentially further breaks $Y_a$ symmetry, and study the corresponding integrable statistical mechanical model. In this case the situation would resemble the link between the golden chain and the one-parameter extension \cite{Fendley_2004}, which corresponds to Baxter's hard square model.

Finally, there are several possibilities for analytic computations at the interface between integrability and fusion categories. For example, it may to possible to prove when an `FF' projector Hamiltonian is boost-integrable by using the pentagon identity. It would also be important to formulate a Bethe ansatz to solve these models. Finally, it would be interesting to make contact with recent the non-invertible $S$-matrix bootstrap programme \cite{Copetti:2024rqj,Copetti:2024dcz}.


\smallskip


\begin{acknowledgments}
\paragraph{Acknowledgements.}

We would like to thank Shota Komatsu, Tristan McLoughlin, and Yuji Tachikawa for useful discussions. We thank Bal{\'a}zs Pozsgay for discussions and collaboration on a related project. MdL was
supported in part by SFI and the Royal Society for funding under grants UF160578, RGF$\backslash$ R1$\backslash$ 181011, RGF$\backslash$8EA$\backslash$180167 and RF$\backslash$ ERE$\backslash$ 210373. MdL is also supported by ERC-2022-CoG - FAIM 101088193. LC was supported by RF$\backslash$ERE$\backslash$210373.

\end{acknowledgments}

\bibliography{HaagerupLetter}

\begin{thebibliography}{55}%
\makeatletter
\providecommand \@ifxundefined [1]{%
 \@ifx{#1\undefined}
}%
\providecommand \@ifnum [1]{%
 \ifnum #1\expandafter \@firstoftwo
 \else \expandafter \@secondoftwo
 \fi
}%
\providecommand \@ifx [1]{%
 \ifx #1\expandafter \@firstoftwo
 \else \expandafter \@secondoftwo
 \fi
}%
\providecommand \natexlab [1]{#1}%
\providecommand \enquote  [1]{``#1''}%
\providecommand \bibnamefont  [1]{#1}%
\providecommand \bibfnamefont [1]{#1}%
\providecommand \citenamefont [1]{#1}%
\providecommand \href@noop [0]{\@secondoftwo}%
\providecommand \href [0]{\begingroup \@sanitize@url \@href}%
\providecommand \@href[1]{\@@startlink{#1}\@@href}%
\providecommand \@@href[1]{\endgroup#1\@@endlink}%
\providecommand \@sanitize@url [0]{\catcode `\\12\catcode `\$12\catcode
  `\&12\catcode `\#12\catcode `\^12\catcode `\_12\catcode `\%12\relax}%
\providecommand \@@startlink[1]{}%
\providecommand \@@endlink[0]{}%
\providecommand \url  [0]{\begingroup\@sanitize@url \@url }%
\providecommand \@url [1]{\endgroup\@href {#1}{\urlprefix }}%
\providecommand \urlprefix  [0]{URL }%
\providecommand \Eprint [0]{\href }%
\providecommand \doibase [0]{http://dx.doi.org/}%
\providecommand \selectlanguage [0]{\@gobble}%
\providecommand \bibinfo  [0]{\@secondoftwo}%
\providecommand \bibfield  [0]{\@secondoftwo}%
\providecommand \translation [1]{[#1]}%
\providecommand \BibitemOpen [0]{}%
\providecommand \bibitemStop [0]{}%
\providecommand \bibitemNoStop [0]{.\EOS\space}%
\providecommand \EOS [0]{\spacefactor3000\relax}%
\providecommand \BibitemShut  [1]{\csname bibitem#1\endcsname}%
\let\auto@bib@innerbib\@empty
\bibitem [{\citenamefont {McGreevy}(2023)}]{McGreevy:2022oyu}%
  \BibitemOpen
  \bibfield  {author} {\bibinfo {author} {\bibfnamefont {John}\ \bibnamefont
  {McGreevy}},\ }\bibfield  {title} {\enquote {\bibinfo {title} {{Generalized
  Symmetries in Condensed Matter}},}\ }\href {\doibase
  10.1146/annurev-conmatphys-040721-021029} {\bibfield  {journal} {\bibinfo
  {journal} {Ann. Rev. Condensed Matter Phys.}\ }\textbf {\bibinfo {volume}
  {14}},\ \bibinfo {pages} {57--82} (\bibinfo {year} {2023})},\ \Eprint
  {http://arxiv.org/abs/2204.03045} {arXiv:2204.03045 [cond-mat.str-el]}
  \BibitemShut {NoStop}%
\bibitem [{\citenamefont {Cordova}\ \emph {et~al.}(2022)\citenamefont
  {Cordova}, \citenamefont {Dumitrescu}, \citenamefont {Intriligator},\ and\
  \citenamefont {Shao}}]{Cordova:2022ruw}%
  \BibitemOpen
  \bibfield  {author} {\bibinfo {author} {\bibfnamefont {Clay}\ \bibnamefont
  {Cordova}}, \bibinfo {author} {\bibfnamefont {Thomas~T.}\ \bibnamefont
  {Dumitrescu}}, \bibinfo {author} {\bibfnamefont {Kenneth}\ \bibnamefont
  {Intriligator}}, \ and\ \bibinfo {author} {\bibfnamefont {Shu-Heng}\
  \bibnamefont {Shao}},\ }\bibfield  {title} {\enquote {\bibinfo {title}
  {{Snowmass White Paper: Generalized Symmetries in Quantum Field Theory and
  Beyond}},}\ }in\ \href@noop {} {\emph {\bibinfo {booktitle} {{Snowmass
  2021}}}}\ (\bibinfo {year} {2022})\ \Eprint {http://arxiv.org/abs/2205.09545}
  {arXiv:2205.09545 [hep-th]} \BibitemShut {NoStop}%
\bibitem [{\citenamefont {Bhardwaj}\ \emph {et~al.}(2024)\citenamefont
  {Bhardwaj}, \citenamefont {Bottini}, \citenamefont {Fraser-Taliente},
  \citenamefont {Gladden}, \citenamefont {Gould}, \citenamefont {Platschorre},\
  and\ \citenamefont {Tillim}}]{Bhardwaj:2023kri}%
  \BibitemOpen
  \bibfield  {author} {\bibinfo {author} {\bibfnamefont {Lakshya}\ \bibnamefont
  {Bhardwaj}}, \bibinfo {author} {\bibfnamefont {Lea~E.}\ \bibnamefont
  {Bottini}}, \bibinfo {author} {\bibfnamefont {Ludovic}\ \bibnamefont
  {Fraser-Taliente}}, \bibinfo {author} {\bibfnamefont {Liam}\ \bibnamefont
  {Gladden}}, \bibinfo {author} {\bibfnamefont {Dewi S.~W.}\ \bibnamefont
  {Gould}}, \bibinfo {author} {\bibfnamefont {Arthur}\ \bibnamefont
  {Platschorre}}, \ and\ \bibinfo {author} {\bibfnamefont {Hannah}\
  \bibnamefont {Tillim}},\ }\bibfield  {title} {\enquote {\bibinfo {title}
  {{Lectures on generalized symmetries}},}\ }\href {\doibase
  10.1016/j.physrep.2023.11.002} {\bibfield  {journal} {\bibinfo  {journal}
  {Phys. Rept.}\ }\textbf {\bibinfo {volume} {1051}},\ \bibinfo {pages} {1--87}
  (\bibinfo {year} {2024})},\ \Eprint {http://arxiv.org/abs/2307.07547}
  {arXiv:2307.07547 [hep-th]} \BibitemShut {NoStop}%
\bibitem [{\citenamefont {Schafer-Nameki}(2024)}]{Schafer-Nameki:2023jdn}%
  \BibitemOpen
  \bibfield  {author} {\bibinfo {author} {\bibfnamefont {Sakura}\ \bibnamefont
  {Schafer-Nameki}},\ }\bibfield  {title} {\enquote {\bibinfo {title} {{ICTP
  lectures on (non-)invertible generalized symmetries}},}\ }\href {\doibase
  10.1016/j.physrep.2024.01.007} {\bibfield  {journal} {\bibinfo  {journal}
  {Phys. Rept.}\ }\textbf {\bibinfo {volume} {1063}},\ \bibinfo {pages} {1--55}
  (\bibinfo {year} {2024})},\ \Eprint {http://arxiv.org/abs/2305.18296}
  {arXiv:2305.18296 [hep-th]} \BibitemShut {NoStop}%
\bibitem [{\citenamefont {Shao}(2023)}]{Shao:2023gho}%
  \BibitemOpen
  \bibfield  {author} {\bibinfo {author} {\bibfnamefont {Shu-Heng}\
  \bibnamefont {Shao}},\ }\bibfield  {title} {\enquote {\bibinfo {title}
  {{What's Done Cannot Be Undone: TASI Lectures on Non-Invertible
  Symmetries}},}\ }\href@noop {} {\  (\bibinfo {year} {2023})},\ \Eprint
  {http://arxiv.org/abs/2308.00747} {arXiv:2308.00747 [hep-th]} \BibitemShut
  {NoStop}%
\bibitem [{\citenamefont {Kramers}\ and\ \citenamefont
  {Wannier}(1941)}]{PhysRev.60.263}%
  \BibitemOpen
  \bibfield  {author} {\bibinfo {author} {\bibfnamefont {H.~A.}\ \bibnamefont
  {Kramers}}\ and\ \bibinfo {author} {\bibfnamefont {G.~H.}\ \bibnamefont
  {Wannier}},\ }\bibfield  {title} {\enquote {\bibinfo {title} {Statistics of
  the two-dimensional ferromagnet. part ii},}\ }\href {\doibase
  10.1103/PhysRev.60.263} {\bibfield  {journal} {\bibinfo  {journal} {Phys.
  Rev.}\ }\textbf {\bibinfo {volume} {60}},\ \bibinfo {pages} {263--276}
  (\bibinfo {year} {1941})}\BibitemShut {NoStop}%
\bibitem [{\citenamefont {Aasen}\ \emph {et~al.}(2016)\citenamefont {Aasen},
  \citenamefont {Mong},\ and\ \citenamefont {Fendley}}]{Aasen:2016dop}%
  \BibitemOpen
  \bibfield  {author} {\bibinfo {author} {\bibfnamefont {David}\ \bibnamefont
  {Aasen}}, \bibinfo {author} {\bibfnamefont {Roger S.~K.}\ \bibnamefont
  {Mong}}, \ and\ \bibinfo {author} {\bibfnamefont {Paul}\ \bibnamefont
  {Fendley}},\ }\bibfield  {title} {\enquote {\bibinfo {title} {{Topological
  Defects on the Lattice I: The Ising model}},}\ }\href {\doibase
  10.1088/1751-8113/49/35/354001} {\bibfield  {journal} {\bibinfo  {journal}
  {J. Phys. A}\ }\textbf {\bibinfo {volume} {49}},\ \bibinfo {pages} {354001}
  (\bibinfo {year} {2016})},\ \Eprint {http://arxiv.org/abs/1601.07185}
  {arXiv:1601.07185 [cond-mat.stat-mech]} \BibitemShut {NoStop}%
\bibitem [{\citenamefont {Aasen}\ \emph {et~al.}(2020)\citenamefont {Aasen},
  \citenamefont {Fendley},\ and\ \citenamefont {Mong}}]{Aasen:2020jwb}%
  \BibitemOpen
  \bibfield  {author} {\bibinfo {author} {\bibfnamefont {David}\ \bibnamefont
  {Aasen}}, \bibinfo {author} {\bibfnamefont {Paul}\ \bibnamefont {Fendley}}, \
  and\ \bibinfo {author} {\bibfnamefont {Roger S.~K.}\ \bibnamefont {Mong}},\
  }\bibfield  {title} {\enquote {\bibinfo {title} {{Topological Defects on the
  Lattice: Dualities and Degeneracies}},}\ }\href@noop {} {\  (\bibinfo {year}
  {2020})},\ \Eprint {http://arxiv.org/abs/2008.08598} {arXiv:2008.08598
  [cond-mat.stat-mech]} \BibitemShut {NoStop}%
\bibitem [{\citenamefont {Inamura}\ and\ \citenamefont
  {Ohmori}(2024)}]{Inamura:2023qzl}%
  \BibitemOpen
  \bibfield  {author} {\bibinfo {author} {\bibfnamefont {Kansei}\ \bibnamefont
  {Inamura}}\ and\ \bibinfo {author} {\bibfnamefont {Kantaro}\ \bibnamefont
  {Ohmori}},\ }\bibfield  {title} {\enquote {\bibinfo {title} {{Fusion surface
  models: 2+1d lattice models from fusion 2-categories}},}\ }\href {\doibase
  10.21468/SciPostPhys.16.6.143} {\bibfield  {journal} {\bibinfo  {journal}
  {SciPost Phys.}\ }\textbf {\bibinfo {volume} {16}},\ \bibinfo {pages} {143}
  (\bibinfo {year} {2024})},\ \Eprint {http://arxiv.org/abs/2305.05774}
  {arXiv:2305.05774 [cond-mat.str-el]} \BibitemShut {NoStop}%
\bibitem [{\citenamefont {Etingof}\ \emph {et~al.}(2015)\citenamefont
  {Etingof}, \citenamefont {Gelaki}, \citenamefont {Nikshych},\ and\
  \citenamefont {Ostrik}}]{fusion}%
  \BibitemOpen
  \bibfield  {author} {\bibinfo {author} {\bibfnamefont {P.}~\bibnamefont
  {Etingof}}, \bibinfo {author} {\bibfnamefont {S.}~\bibnamefont {Gelaki}},
  \bibinfo {author} {\bibfnamefont {D.}~\bibnamefont {Nikshych}}, \ and\
  \bibinfo {author} {\bibfnamefont {V}~\bibnamefont {Ostrik}},\ }\href@noop {}
  {\emph {\bibinfo {title} {Tensor Categories}}}\ (\bibinfo  {publisher}
  {American Mathematical Society},\ \bibinfo {year} {2015})\BibitemShut
  {NoStop}%
\bibitem [{\citenamefont {Feiguin}\ \emph {et~al.}(2007)\citenamefont
  {Feiguin}, \citenamefont {Trebst}, \citenamefont {Ludwig}, \citenamefont
  {Troyer}, \citenamefont {Kitaev}, \citenamefont {Wang},\ and\ \citenamefont
  {Freedman}}]{Feiguin:2006ydp}%
  \BibitemOpen
  \bibfield  {author} {\bibinfo {author} {\bibfnamefont {Adrian}\ \bibnamefont
  {Feiguin}}, \bibinfo {author} {\bibfnamefont {Simon}\ \bibnamefont {Trebst}},
  \bibinfo {author} {\bibfnamefont {Andreas W.~W.}\ \bibnamefont {Ludwig}},
  \bibinfo {author} {\bibfnamefont {Matthias}\ \bibnamefont {Troyer}}, \bibinfo
  {author} {\bibfnamefont {Alexei}\ \bibnamefont {Kitaev}}, \bibinfo {author}
  {\bibfnamefont {Zhenghan}\ \bibnamefont {Wang}}, \ and\ \bibinfo {author}
  {\bibfnamefont {Michael~H.}\ \bibnamefont {Freedman}},\ }\bibfield  {title}
  {\enquote {\bibinfo {title} {{Interacting anyons in topological quantum
  liquids: The golden chain}},}\ }\href {\doibase
  10.1103/PhysRevLett.98.160409} {\bibfield  {journal} {\bibinfo  {journal}
  {Phys. Rev. Lett.}\ }\textbf {\bibinfo {volume} {98}},\ \bibinfo {pages}
  {160409} (\bibinfo {year} {2007})},\ \Eprint
  {http://arxiv.org/abs/cond-mat/0612341} {arXiv:cond-mat/0612341} \BibitemShut
  {NoStop}%
\bibitem [{\citenamefont {Trebst}\ \emph {et~al.}(2008)\citenamefont {Trebst},
  \citenamefont {Troyer}, \citenamefont {Wang},\ and\ \citenamefont
  {Ludwig}}]{Trebst_2008}%
  \BibitemOpen
  \bibfield  {author} {\bibinfo {author} {\bibfnamefont {Simon}\ \bibnamefont
  {Trebst}}, \bibinfo {author} {\bibfnamefont {Matthias}\ \bibnamefont
  {Troyer}}, \bibinfo {author} {\bibfnamefont {Zhenghan}\ \bibnamefont {Wang}},
  \ and\ \bibinfo {author} {\bibfnamefont {Andreas W.~W.}\ \bibnamefont
  {Ludwig}},\ }\bibfield  {title} {\enquote {\bibinfo {title} {A short
  introduction to fibonacci anyon models},}\ }\href {\doibase
  10.1143/ptps.176.384} {\bibfield  {journal} {\bibinfo  {journal} {Progress of
  Theoretical Physics Supplement}\ }\textbf {\bibinfo {volume} {176}},\
  \bibinfo {pages} {384–407} (\bibinfo {year} {2008})}\BibitemShut {NoStop}%
\bibitem [{\citenamefont {Fendley}\ \emph {et~al.}(2004)\citenamefont
  {Fendley}, \citenamefont {Sengupta},\ and\ \citenamefont
  {Sachdev}}]{Fendley_2004}%
  \BibitemOpen
  \bibfield  {author} {\bibinfo {author} {\bibfnamefont {Paul}\ \bibnamefont
  {Fendley}}, \bibinfo {author} {\bibfnamefont {K.}~\bibnamefont {Sengupta}}, \
  and\ \bibinfo {author} {\bibfnamefont {Subir}\ \bibnamefont {Sachdev}},\
  }\bibfield  {title} {\enquote {\bibinfo {title} {Competing density-wave
  orders in a one-dimensional hard-boson model},}\ }\href {\doibase
  10.1103/physrevb.69.075106} {\bibfield  {journal} {\bibinfo  {journal}
  {Physical Review B}\ }\textbf {\bibinfo {volume} {69}} (\bibinfo {year}
  {2004}),\ 10.1103/physrevb.69.075106}\BibitemShut {NoStop}%
\bibitem [{\citenamefont {Baxter}(1982)}]{Baxter-Book}%
  \BibitemOpen
  \bibfield  {author} {\bibinfo {author} {\bibfnamefont {R.~J.}\ \bibnamefont
  {Baxter}},\ }\href@noop {} {\emph {\bibinfo {title} {Exactly solved models in
  statistical mechanics}}}\ (\bibinfo  {publisher} {London: Academic Press
  Inc},\ \bibinfo {year} {1982})\BibitemShut {NoStop}%
\bibitem [{\citenamefont {Andrews}\ \emph {et~al.}(1984)\citenamefont
  {Andrews}, \citenamefont {Baxter},\ and\ \citenamefont {Forrester}}]{RSOS-1}%
  \BibitemOpen
  \bibfield  {author} {\bibinfo {author} {\bibfnamefont {G.~E.}\ \bibnamefont
  {Andrews}}, \bibinfo {author} {\bibfnamefont {R.~J.}\ \bibnamefont {Baxter}},
  \ and\ \bibinfo {author} {\bibfnamefont {P.~J.}\ \bibnamefont {Forrester}},\
  }\bibfield  {title} {\enquote {\bibinfo {title} {Eight-vertex sos model and
  generalized rogers-ramanujan-type identities},}\ }\href {\doibase
  10.1007/BF01014383} {\bibfield  {journal} {\bibinfo  {journal} {J. Stat.
  Phys.}\ }\textbf {\bibinfo {volume} {35}},\ \bibinfo {pages} {193} (\bibinfo
  {year} {1984})}\BibitemShut {NoStop}%
\bibitem [{\citenamefont {Baxter}\ and\ \citenamefont
  {Forrester}(1985)}]{RSOS-2}%
  \BibitemOpen
  \bibfield  {author} {\bibinfo {author} {\bibfnamefont {R.~J.}\ \bibnamefont
  {Baxter}}\ and\ \bibinfo {author} {\bibfnamefont {P.~J.}\ \bibnamefont
  {Forrester}},\ }\bibfield  {title} {\enquote {\bibinfo {title} {Further exact
  solutions of the eight-vertex sos model and generalizations of the
  rogers-ramanujan identities},}\ }\href {\doibase 10.1007/BF01010471}
  {\bibfield  {journal} {\bibinfo  {journal} {J. Stat. Phys.}\ }\textbf
  {\bibinfo {volume} {38}},\ \bibinfo {pages} {435} (\bibinfo {year}
  {1985})}\BibitemShut {NoStop}%
\bibitem [{\citenamefont {{Bianchini}}\ \emph {et~al.}(2015)\citenamefont
  {{Bianchini}}, \citenamefont {{Ercolessi}}, \citenamefont {{Pearce}},\ and\
  \citenamefont {{Ravanini}}}]{RSOS-H}%
  \BibitemOpen
  \bibfield  {author} {\bibinfo {author} {\bibfnamefont {Davide}\ \bibnamefont
  {{Bianchini}}}, \bibinfo {author} {\bibfnamefont {Elisa}\ \bibnamefont
  {{Ercolessi}}}, \bibinfo {author} {\bibfnamefont {Paul~A.}\ \bibnamefont
  {{Pearce}}}, \ and\ \bibinfo {author} {\bibfnamefont {Francesco}\
  \bibnamefont {{Ravanini}}},\ }\bibfield  {title} {\enquote {\bibinfo {title}
  {Rsos quantum chains associated with off-critical minimal models and $z_n$
  parafermions},}\ }\href {\doibase 10.1088/1742-5468/2015/03/P03010}
  {\bibfield  {journal} {\bibinfo  {journal} {J. Stat. Mech.}\ }\textbf
  {\bibinfo {volume} {2015}},\ \bibinfo {eid} {03010} (\bibinfo {year}
  {2015})},\ \Eprint {http://arxiv.org/abs/1412.4942} {arXiv:1412.4942
  [hep-th]} \BibitemShut {NoStop}%
\bibitem [{\citenamefont {Urban}\ \emph {et~al.}(2009)\citenamefont {Urban},
  \citenamefont {Johnson}, \citenamefont {Henage}, \citenamefont {Isenhower},
  \citenamefont {Yavuz}, \citenamefont {Walker},\ and\ \citenamefont
  {Saffman}}]{Urban_2009}%
  \BibitemOpen
  \bibfield  {author} {\bibinfo {author} {\bibfnamefont {E.}~\bibnamefont
  {Urban}}, \bibinfo {author} {\bibfnamefont {T.~A.}\ \bibnamefont {Johnson}},
  \bibinfo {author} {\bibfnamefont {T.}~\bibnamefont {Henage}}, \bibinfo
  {author} {\bibfnamefont {L.}~\bibnamefont {Isenhower}}, \bibinfo {author}
  {\bibfnamefont {D.~D.}\ \bibnamefont {Yavuz}}, \bibinfo {author}
  {\bibfnamefont {T.~G.}\ \bibnamefont {Walker}}, \ and\ \bibinfo {author}
  {\bibfnamefont {M.}~\bibnamefont {Saffman}},\ }\bibfield  {title} {\enquote
  {\bibinfo {title} {Observation of rydberg blockade between two atoms},}\
  }\href {\doibase 10.1038/nphys1178} {\bibfield  {journal} {\bibinfo
  {journal} {Nature Physics}\ }\textbf {\bibinfo {volume} {5}},\ \bibinfo
  {pages} {110–114} (\bibinfo {year} {2009})}\BibitemShut {NoStop}%
\bibitem [{\citenamefont {Lesanovsky}\ and\ \citenamefont
  {Katsura}(2012)}]{PhysRevA.86.041601}%
  \BibitemOpen
  \bibfield  {author} {\bibinfo {author} {\bibfnamefont {Igor}\ \bibnamefont
  {Lesanovsky}}\ and\ \bibinfo {author} {\bibfnamefont {Hosho}\ \bibnamefont
  {Katsura}},\ }\bibfield  {title} {\enquote {\bibinfo {title} {Interacting
  fibonacci anyons in a rydberg gas},}\ }\href {\doibase
  10.1103/PhysRevA.86.041601} {\bibfield  {journal} {\bibinfo  {journal} {Phys.
  Rev. A}\ }\textbf {\bibinfo {volume} {86}},\ \bibinfo {pages} {041601}
  (\bibinfo {year} {2012})}\BibitemShut {NoStop}%
\bibitem [{\citenamefont {Serbyn}\ \emph {et~al.}(2021)\citenamefont {Serbyn},
  \citenamefont {Abanin},\ and\ \citenamefont {Papi\'c}}]{Serbyn:2020wys}%
  \BibitemOpen
  \bibfield  {author} {\bibinfo {author} {\bibfnamefont {Maksym}\ \bibnamefont
  {Serbyn}}, \bibinfo {author} {\bibfnamefont {Dmitry~A.}\ \bibnamefont
  {Abanin}}, \ and\ \bibinfo {author} {\bibfnamefont {Zlatko}\ \bibnamefont
  {Papi\'c}},\ }\bibfield  {title} {\enquote {\bibinfo {title} {{Quantum
  many-body scars and weak breaking of ergodicity}},}\ }\href {\doibase
  10.1038/s41567-021-01230-2} {\bibfield  {journal} {\bibinfo  {journal}
  {Nature Phys.}\ }\textbf {\bibinfo {volume} {17}},\ \bibinfo {pages}
  {675--685} (\bibinfo {year} {2021})},\ \Eprint
  {http://arxiv.org/abs/2011.09486} {arXiv:2011.09486 [quant-ph]} \BibitemShut
  {NoStop}%
\bibitem [{\citenamefont {Turner}\ \emph {et~al.}(2018)\citenamefont {Turner},
  \citenamefont {Michailidis}, \citenamefont {Abanin}, \citenamefont {Serbyn},\
  and\ \citenamefont {Papi\ifmmode~\acute{c}\else
  \'{c}\fi{}}}]{PhysRevB.98.155134}%
  \BibitemOpen
  \bibfield  {author} {\bibinfo {author} {\bibfnamefont {C.~J.}\ \bibnamefont
  {Turner}}, \bibinfo {author} {\bibfnamefont {A.~A.}\ \bibnamefont
  {Michailidis}}, \bibinfo {author} {\bibfnamefont {D.~A.}\ \bibnamefont
  {Abanin}}, \bibinfo {author} {\bibfnamefont {M.}~\bibnamefont {Serbyn}}, \
  and\ \bibinfo {author} {\bibfnamefont {Z.}~\bibnamefont
  {Papi\ifmmode~\acute{c}\else \'{c}\fi{}}},\ }\bibfield  {title} {\enquote
  {\bibinfo {title} {Quantum scarred eigenstates in a rydberg atom chain:
  Entanglement, breakdown of thermalization, and stability to perturbations},}\
  }\href {\doibase 10.1103/PhysRevB.98.155134} {\bibfield  {journal} {\bibinfo
  {journal} {Phys. Rev. B}\ }\textbf {\bibinfo {volume} {98}},\ \bibinfo
  {pages} {155134} (\bibinfo {year} {2018})}\BibitemShut {NoStop}%
\bibitem [{\citenamefont {Lesanovsky}(2011)}]{lesanovsky}%
  \BibitemOpen
  \bibfield  {author} {\bibinfo {author} {\bibfnamefont {Igor}\ \bibnamefont
  {Lesanovsky}},\ }\bibfield  {title} {\enquote {\bibinfo {title} {Many-body
  spin interactions and the ground state of a dense rydberg lattice gas},}\
  }\href {\doibase 10.1103/PhysRevLett.106.025301} {\bibfield  {journal}
  {\bibinfo  {journal} {Phys. Rev. Lett.}\ }\textbf {\bibinfo {volume} {106}},\
  \bibinfo {pages} {025301} (\bibinfo {year} {2011})}\BibitemShut {NoStop}%
\bibitem [{\citenamefont {{Lesanovsky}}(2012)}]{lesanovsky-model}%
  \BibitemOpen
  \bibfield  {author} {\bibinfo {author} {\bibfnamefont {Igor}\ \bibnamefont
  {{Lesanovsky}}},\ }\bibfield  {title} {\enquote {\bibinfo {title} {{Liquid
  Ground State, Gap, and Excited States of a Strongly Correlated Spin
  Chain}},}\ }\href {\doibase 10.1103/PhysRevLett.108.105301} {\bibfield
  {journal} {\bibinfo  {journal} {Phys. Rev. Lett.}\ }\textbf {\bibinfo
  {volume} {108}},\ \bibinfo {eid} {105301} (\bibinfo {year} {2012})},\ \Eprint
  {http://arxiv.org/abs/1110.1815} {arXiv:1110.1815 [cond-mat.str-el]}
  \BibitemShut {NoStop}%
\bibitem [{\citenamefont {{Mark}}\ \emph {et~al.}(2020)\citenamefont {{Mark}},
  \citenamefont {{Lin}},\ and\ \citenamefont {{Motrunich}}}]{pxp-int}%
  \BibitemOpen
  \bibfield  {author} {\bibinfo {author} {\bibfnamefont {Daniel~K.}\
  \bibnamefont {{Mark}}}, \bibinfo {author} {\bibfnamefont {Cheng-Ju}\
  \bibnamefont {{Lin}}}, \ and\ \bibinfo {author} {\bibfnamefont {Olexei~I.}\
  \bibnamefont {{Motrunich}}},\ }\bibfield  {title} {\enquote {\bibinfo {title}
  {{Exact eigenstates in the Lesanovsky model, proximity to integrability and
  the PXP model, and approximate scar states}},}\ }\href {\doibase
  10.1103/PhysRevB.101.094308} {\bibfield  {journal} {\bibinfo  {journal}
  {Phys. Rev. B}\ }\textbf {\bibinfo {volume} {101}},\ \bibinfo {eid} {094308}
  (\bibinfo {year} {2020})},\ \Eprint {http://arxiv.org/abs/1911.11305}
  {arXiv:1911.11305 [cond-mat.quant-gas]} \BibitemShut {NoStop}%
\bibitem [{\citenamefont {{Alcaraz}}\ and\ \citenamefont
  {{Bariev}}(1999{\natexlab{a}})}]{constrained1}%
  \BibitemOpen
  \bibfield  {author} {\bibinfo {author} {\bibfnamefont {F.~C.}\ \bibnamefont
  {{Alcaraz}}}\ and\ \bibinfo {author} {\bibfnamefont {R.~Z.}\ \bibnamefont
  {{Bariev}}},\ }\bibfield  {title} {\enquote {\bibinfo {title} {{An Exactly
  Solvable Constrained XXZ Chain}},}\ }\href@noop {} {\bibfield  {journal}
  {\bibinfo  {journal} {arXiv e-prints}\ } (\bibinfo {year}
  {1999}{\natexlab{a}})},\ \Eprint {http://arxiv.org/abs/cond-mat/9904042}
  {arXiv:cond-mat/9904042 [cond-mat.stat-mech]} \BibitemShut {NoStop}%
\bibitem [{\citenamefont {{Alcaraz}}\ and\ \citenamefont
  {{Bariev}}(1999{\natexlab{b}})}]{constrained2}%
  \BibitemOpen
  \bibfield  {author} {\bibinfo {author} {\bibfnamefont {F.~C.}\ \bibnamefont
  {{Alcaraz}}}\ and\ \bibinfo {author} {\bibfnamefont {R.~Z.}\ \bibnamefont
  {{Bariev}}},\ }\bibfield  {title} {\enquote {\bibinfo {title} {{Integrable
  models of strongly correlated particles with correlated hopping}},}\ }\href
  {\doibase 10.1103/PhysRevB.59.3373} {\bibfield  {journal} {\bibinfo
  {journal} {Phys. Rev. B}\ }\textbf {\bibinfo {volume} {59}},\ \bibinfo
  {pages} {3373--3376} (\bibinfo {year} {1999}{\natexlab{b}})},\ \Eprint
  {http://arxiv.org/abs/cond-mat/9904129} {arXiv:cond-mat/9904129
  [cond-mat.stat-mech]} \BibitemShut {NoStop}%
\bibitem [{\citenamefont {Haagerup}(1993)}]{haagerup}%
  \BibitemOpen
  \bibfield  {author} {\bibinfo {author} {\bibfnamefont {Uffe}\ \bibnamefont
  {Haagerup}},\ }\bibfield  {title} {\enquote {\bibinfo {title} {{Principal
  graphs of subfactors in the index range $4 < [M : N] < 3 + \sqrt{2}$}},}\
  }\href@noop {} {\bibfield  {journal} {\bibinfo  {journal} {Subfactors
  (Proceedings of Taniguchi Symposium, Kyuzeso)}\ } (\bibinfo {year}
  {1993})}\BibitemShut {NoStop}%
\bibitem [{\citenamefont {Grossman}\ and\ \citenamefont
  {Snyder}(2012)}]{Grossman_2012}%
  \BibitemOpen
  \bibfield  {author} {\bibinfo {author} {\bibfnamefont {Pinhas}\ \bibnamefont
  {Grossman}}\ and\ \bibinfo {author} {\bibfnamefont {Noah}\ \bibnamefont
  {Snyder}},\ }\bibfield  {title} {\enquote {\bibinfo {title} {Quantum
  subgroups of the haagerup fusion categories},}\ }\href {\doibase
  10.1007/s00220-012-1427-x} {\bibfield  {journal} {\bibinfo  {journal}
  {Communications in Mathematical Physics}\ }\textbf {\bibinfo {volume}
  {311}},\ \bibinfo {pages} {617–643} (\bibinfo {year} {2012})}\BibitemShut
  {NoStop}%
\bibitem [{\citenamefont {Osborne}\ \emph {et~al.}(2019)\citenamefont
  {Osborne}, \citenamefont {Stiegemann},\ and\ \citenamefont
  {Wolf}}]{osborne2019fsymbols}%
  \BibitemOpen
  \bibfield  {author} {\bibinfo {author} {\bibfnamefont {Tobias~J.}\
  \bibnamefont {Osborne}}, \bibinfo {author} {\bibfnamefont {Deniz~E.}\
  \bibnamefont {Stiegemann}}, \ and\ \bibinfo {author} {\bibfnamefont {Ramona}\
  \bibnamefont {Wolf}},\ }\href@noop {} {\enquote {\bibinfo {title} {The
  f-symbols for the h3 fusion category},}\ } (\bibinfo {year} {2019}),\ \Eprint
  {http://arxiv.org/abs/1906.01322} {arXiv:1906.01322} \BibitemShut {NoStop}%
\bibitem [{\citenamefont {Huang}\ and\ \citenamefont
  {Lin}(2020)}]{Huang:2020lox}%
  \BibitemOpen
  \bibfield  {author} {\bibinfo {author} {\bibfnamefont {Tzu-Chen}\
  \bibnamefont {Huang}}\ and\ \bibinfo {author} {\bibfnamefont {Ying-Hsuan}\
  \bibnamefont {Lin}},\ }\bibfield  {title} {\enquote {\bibinfo {title} {{The
  $F$-Symbols for Transparent Haagerup-Izumi Categories with $G =
  \mathbb{Z}_{2n+1}$}},}\ }\href@noop {} {\  (\bibinfo {year} {2020})},\
  \Eprint {http://arxiv.org/abs/2007.00670} {arXiv:2007.00670 [math.CT]}
  \BibitemShut {NoStop}%
\bibitem [{\citenamefont {Wolf}(2020)}]{Wolf:2020qdo}%
  \BibitemOpen
  \bibfield  {author} {\bibinfo {author} {\bibfnamefont {Ramona}\ \bibnamefont
  {Wolf}},\ }\emph {\bibinfo {title} {{Microscopic Models for Fusion
  Categories}}},\ \href {\doibase 10.15488/10324} {Ph.D. thesis},\ \bibinfo
  {school} {Leibniz U., Hannover} (\bibinfo {year} {2020}),\ \Eprint
  {http://arxiv.org/abs/2101.04154} {arXiv:2101.04154 [math-ph]} \BibitemShut
  {NoStop}%
\bibitem [{\citenamefont {Huang}\ \emph {et~al.}(2022)\citenamefont {Huang},
  \citenamefont {Lin}, \citenamefont {Ohmori}, \citenamefont {Tachikawa},\ and\
  \citenamefont {Tezuka}}]{Huang:2021nvb}%
  \BibitemOpen
  \bibfield  {author} {\bibinfo {author} {\bibfnamefont {Tzu-Chen}\
  \bibnamefont {Huang}}, \bibinfo {author} {\bibfnamefont {Ying-Hsuan}\
  \bibnamefont {Lin}}, \bibinfo {author} {\bibfnamefont {Kantaro}\ \bibnamefont
  {Ohmori}}, \bibinfo {author} {\bibfnamefont {Yuji}\ \bibnamefont
  {Tachikawa}}, \ and\ \bibinfo {author} {\bibfnamefont {Masaki}\ \bibnamefont
  {Tezuka}},\ }\bibfield  {title} {\enquote {\bibinfo {title} {{Numerical
  Evidence for a Haagerup Conformal Field Theory}},}\ }\href {\doibase
  10.1103/PhysRevLett.128.231603} {\bibfield  {journal} {\bibinfo  {journal}
  {Phys. Rev. Lett.}\ }\textbf {\bibinfo {volume} {128}},\ \bibinfo {pages}
  {231603} (\bibinfo {year} {2022})},\ \Eprint
  {http://arxiv.org/abs/2110.03008} {arXiv:2110.03008 [cond-mat.stat-mech]}
  \BibitemShut {NoStop}%
\bibitem [{\citenamefont {Vanhove}\ \emph {et~al.}(2022)\citenamefont
  {Vanhove}, \citenamefont {Lootens}, \citenamefont {Van~Damme}, \citenamefont
  {Wolf}, \citenamefont {Osborne}, \citenamefont {Haegeman},\ and\
  \citenamefont {Verstraete}}]{Vanhove:2021zop}%
  \BibitemOpen
  \bibfield  {author} {\bibinfo {author} {\bibfnamefont {Robijn}\ \bibnamefont
  {Vanhove}}, \bibinfo {author} {\bibfnamefont {Laurens}\ \bibnamefont
  {Lootens}}, \bibinfo {author} {\bibfnamefont {Maarten}\ \bibnamefont
  {Van~Damme}}, \bibinfo {author} {\bibfnamefont {Ramona}\ \bibnamefont
  {Wolf}}, \bibinfo {author} {\bibfnamefont {Tobias~J.}\ \bibnamefont
  {Osborne}}, \bibinfo {author} {\bibfnamefont {Jutho}\ \bibnamefont
  {Haegeman}}, \ and\ \bibinfo {author} {\bibfnamefont {Frank}\ \bibnamefont
  {Verstraete}},\ }\bibfield  {title} {\enquote {\bibinfo {title} {{Critical
  Lattice Model for a Haagerup Conformal Field Theory}},}\ }\href {\doibase
  10.1103/PhysRevLett.128.231602} {\bibfield  {journal} {\bibinfo  {journal}
  {Phys. Rev. Lett.}\ }\textbf {\bibinfo {volume} {128}},\ \bibinfo {pages}
  {231602} (\bibinfo {year} {2022})},\ \Eprint
  {http://arxiv.org/abs/2110.03532} {arXiv:2110.03532 [cond-mat.stat-mech]}
  \BibitemShut {NoStop}%
\bibitem [{\citenamefont {de~Leeuw}\ \emph {et~al.}(2019)\citenamefont
  {de~Leeuw}, \citenamefont {Pribytok},\ and\ \citenamefont
  {Ryan}}]{marius-classification-1}%
  \BibitemOpen
  \bibfield  {author} {\bibinfo {author} {\bibfnamefont {Marius}\ \bibnamefont
  {de~Leeuw}}, \bibinfo {author} {\bibfnamefont {Anton}\ \bibnamefont
  {Pribytok}}, \ and\ \bibinfo {author} {\bibfnamefont {Paul}\ \bibnamefont
  {Ryan}},\ }\bibfield  {title} {\enquote {\bibinfo {title} {{Classifying
  two-dimensional integrable spin chains}},}\ }\href {\doibase
  10.1088/1751-8121/ab529f} {\bibfield  {journal} {\bibinfo  {journal} {J.
  Phys. A}\ }\textbf {\bibinfo {volume} {52}},\ \bibinfo {pages} {505201}
  (\bibinfo {year} {2019})},\ \Eprint {http://arxiv.org/abs/1904.12005}
  {arXiv:1904.12005 [math-ph]} \BibitemShut {NoStop}%
\bibitem [{\citenamefont {de~Leeuw}\ \emph
  {et~al.}(2020{\natexlab{a}})\citenamefont {de~Leeuw}, \citenamefont
  {Paletta}, \citenamefont {Pribytok}, \citenamefont {Retore},\ and\
  \citenamefont {Ryan}}]{marius-classification-2}%
  \BibitemOpen
  \bibfield  {author} {\bibinfo {author} {\bibfnamefont {Marius}\ \bibnamefont
  {de~Leeuw}}, \bibinfo {author} {\bibfnamefont {Chiara}\ \bibnamefont
  {Paletta}}, \bibinfo {author} {\bibfnamefont {Anton}\ \bibnamefont
  {Pribytok}}, \bibinfo {author} {\bibfnamefont {Ana}\ \bibnamefont {Retore}},
  \ and\ \bibinfo {author} {\bibfnamefont {Paul}\ \bibnamefont {Ryan}},\
  }\bibfield  {title} {\enquote {\bibinfo {title} {{Classifying
  Nearest-Neighbor Interactions and Deformations of AdS}},}\ }\href {\doibase
  10.1103/PhysRevLett.125.031604} {\bibfield  {journal} {\bibinfo  {journal}
  {Phys. Rev. Lett.}\ }\textbf {\bibinfo {volume} {125}},\ \bibinfo {pages}
  {031604} (\bibinfo {year} {2020}{\natexlab{a}})},\ \Eprint
  {http://arxiv.org/abs/2003.04332} {arXiv:2003.04332 [hep-th]} \BibitemShut
  {NoStop}%
\bibitem [{\citenamefont {de~Leeuw}\ \emph
  {et~al.}(2020{\natexlab{b}})\citenamefont {de~Leeuw}, \citenamefont
  {Pribytok}, \citenamefont {Retore},\ and\ \citenamefont
  {Ryan}}]{marius-classification-3}%
  \BibitemOpen
  \bibfield  {author} {\bibinfo {author} {\bibfnamefont {Marius}\ \bibnamefont
  {de~Leeuw}}, \bibinfo {author} {\bibfnamefont {Anton}\ \bibnamefont
  {Pribytok}}, \bibinfo {author} {\bibfnamefont {Ana}\ \bibnamefont {Retore}},
  \ and\ \bibinfo {author} {\bibfnamefont {Paul}\ \bibnamefont {Ryan}},\
  }\bibfield  {title} {\enquote {\bibinfo {title} {{New integrable 1D models of
  superconductivity}},}\ }\href {\doibase 10.1088/1751-8121/aba860} {\bibfield
  {journal} {\bibinfo  {journal} {J. Phys. A}\ }\textbf {\bibinfo {volume}
  {53}},\ \bibinfo {pages} {385201} (\bibinfo {year} {2020}{\natexlab{b}})},\
  \Eprint {http://arxiv.org/abs/1911.01439} {arXiv:1911.01439 [math-ph]}
  \BibitemShut {NoStop}%
\bibitem [{\citenamefont {de~Leeuw}\ \emph {et~al.}(2021)\citenamefont
  {de~Leeuw}, \citenamefont {Paletta}, \citenamefont {Pribytok}, \citenamefont
  {Retore},\ and\ \citenamefont {Ryan}}]{marius-classification-4}%
  \BibitemOpen
  \bibfield  {author} {\bibinfo {author} {\bibfnamefont {Marius}\ \bibnamefont
  {de~Leeuw}}, \bibinfo {author} {\bibfnamefont {Chiara}\ \bibnamefont
  {Paletta}}, \bibinfo {author} {\bibfnamefont {Anton}\ \bibnamefont
  {Pribytok}}, \bibinfo {author} {\bibfnamefont {Ana}\ \bibnamefont {Retore}},
  \ and\ \bibinfo {author} {\bibfnamefont {Paul}\ \bibnamefont {Ryan}},\
  }\bibfield  {title} {\enquote {\bibinfo {title} {{Yang-Baxter and the Boost:
  splitting the difference}},}\ }\href {\doibase 10.21468/SciPostPhys.11.3.069}
  {\bibfield  {journal} {\bibinfo  {journal} {SciPost Phys.}\ }\textbf
  {\bibinfo {volume} {11}},\ \bibinfo {pages} {069} (\bibinfo {year} {2021})},\
  \Eprint {http://arxiv.org/abs/2010.11231} {arXiv:2010.11231 [math-ph]}
  \BibitemShut {NoStop}%
\bibitem [{\citenamefont {De~Leeuw}\ \emph {et~al.}(2021)\citenamefont
  {De~Leeuw}, \citenamefont {Paletta}, \citenamefont {Pribytok}, \citenamefont
  {Retore},\ and\ \citenamefont {Torrielli}}]{marius-classification-5}%
  \BibitemOpen
  \bibfield  {author} {\bibinfo {author} {\bibfnamefont {Marius}\ \bibnamefont
  {De~Leeuw}}, \bibinfo {author} {\bibfnamefont {Chiara}\ \bibnamefont
  {Paletta}}, \bibinfo {author} {\bibfnamefont {Anton}\ \bibnamefont
  {Pribytok}}, \bibinfo {author} {\bibfnamefont {Ana}\ \bibnamefont {Retore}},
  \ and\ \bibinfo {author} {\bibfnamefont {Alessandro}\ \bibnamefont
  {Torrielli}},\ }\bibfield  {title} {\enquote {\bibinfo {title} {{Free
  Fermions, vertex Hamiltonians, and lower-dimensional AdS/CFT}},}\ }\href
  {\doibase 10.1007/JHEP02(2021)191} {\bibfield  {journal} {\bibinfo  {journal}
  {JHEP}\ }\textbf {\bibinfo {volume} {02}},\ \bibinfo {pages} {191} (\bibinfo
  {year} {2021})},\ \Eprint {http://arxiv.org/abs/2011.08217} {arXiv:2011.08217
  [hep-th]} \BibitemShut {NoStop}%
\bibitem [{\citenamefont {Corcoran}\ and\ \citenamefont
  {de~Leeuw}(2023)}]{marius-classification-6}%
  \BibitemOpen
  \bibfield  {author} {\bibinfo {author} {\bibfnamefont {Luke}\ \bibnamefont
  {Corcoran}}\ and\ \bibinfo {author} {\bibfnamefont {Marius}\ \bibnamefont
  {de~Leeuw}},\ }\bibfield  {title} {\enquote {\bibinfo {title} {{All regular
  $4 \times 4$ solutions of the Yang-Baxter equation}},}\ }\href@noop {} {\
  (\bibinfo {year} {2023})},\ \Eprint {http://arxiv.org/abs/2306.10423}
  {arXiv:2306.10423 [hep-th]} \BibitemShut {NoStop}%
\bibitem [{\citenamefont {Gombor}\ and\ \citenamefont
  {Pozsgay}(2021)}]{Gombor:2021nhn}%
  \BibitemOpen
  \bibfield  {author} {\bibinfo {author} {\bibfnamefont {Tam\'as}\ \bibnamefont
  {Gombor}}\ and\ \bibinfo {author} {\bibfnamefont {Bal\'azs}\ \bibnamefont
  {Pozsgay}},\ }\bibfield  {title} {\enquote {\bibinfo {title} {{Integrable
  spin chains and cellular automata with medium-range interaction}},}\ }\href
  {\doibase 10.1103/PhysRevE.104.054123} {\bibfield  {journal} {\bibinfo
  {journal} {Phys. Rev. E}\ }\textbf {\bibinfo {volume} {104}},\ \bibinfo
  {pages} {054123} (\bibinfo {year} {2021})},\ \Eprint
  {http://arxiv.org/abs/2108.02053} {arXiv:2108.02053 [nlin.SI]} \BibitemShut
  {NoStop}%
\bibitem [{\citenamefont {Corcoran}\ \emph {et~al.}(2024)\citenamefont
  {Corcoran}, \citenamefont {de~Leeuw},\ and\ \citenamefont
  {Pozsgay}}]{Corcoran:2024ofo}%
  \BibitemOpen
  \bibfield  {author} {\bibinfo {author} {\bibfnamefont {Luke}\ \bibnamefont
  {Corcoran}}, \bibinfo {author} {\bibfnamefont {Marius}\ \bibnamefont
  {de~Leeuw}}, \ and\ \bibinfo {author} {\bibfnamefont {Bal\'azs}\ \bibnamefont
  {Pozsgay}},\ }\bibfield  {title} {\enquote {\bibinfo {title} {{Integrable
  models on Rydberg atom chains}},}\ }\href@noop {} {\  (\bibinfo {year}
  {2024})},\ \Eprint {http://arxiv.org/abs/2405.15848} {arXiv:2405.15848
  [cond-mat.str-el]} \BibitemShut {NoStop}%
\bibitem [{\citenamefont {Gils}\ \emph {et~al.}(2013)\citenamefont {Gils},
  \citenamefont {Ardonne}, \citenamefont {Trebst}, \citenamefont {Huse},
  \citenamefont {Ludwig}, \citenamefont {Troyer},\ and\ \citenamefont
  {Wang}}]{Gils_2013}%
  \BibitemOpen
  \bibfield  {author} {\bibinfo {author} {\bibfnamefont {C.}~\bibnamefont
  {Gils}}, \bibinfo {author} {\bibfnamefont {E.}~\bibnamefont {Ardonne}},
  \bibinfo {author} {\bibfnamefont {S.}~\bibnamefont {Trebst}}, \bibinfo
  {author} {\bibfnamefont {D.~A.}\ \bibnamefont {Huse}}, \bibinfo {author}
  {\bibfnamefont {A.~W.~W.}\ \bibnamefont {Ludwig}}, \bibinfo {author}
  {\bibfnamefont {M.}~\bibnamefont {Troyer}}, \ and\ \bibinfo {author}
  {\bibfnamefont {Z.}~\bibnamefont {Wang}},\ }\bibfield  {title} {\enquote
  {\bibinfo {title} {Anyonic quantum spin chains: Spin-1 generalizations and
  topological stability},}\ }\href {\doibase 10.1103/physrevb.87.235120}
  {\bibfield  {journal} {\bibinfo  {journal} {Physical Review B}\ }\textbf
  {\bibinfo {volume} {87}} (\bibinfo {year} {2013}),\
  10.1103/physrevb.87.235120}\BibitemShut {NoStop}%
\bibitem [{\citenamefont {Buican}\ and\ \citenamefont
  {Gromov}(2017)}]{Buican:2017rxc}%
  \BibitemOpen
  \bibfield  {author} {\bibinfo {author} {\bibfnamefont {Matthew}\ \bibnamefont
  {Buican}}\ and\ \bibinfo {author} {\bibfnamefont {Andrey}\ \bibnamefont
  {Gromov}},\ }\bibfield  {title} {\enquote {\bibinfo {title} {{Anyonic Chains,
  Topological Defects, and Conformal Field Theory}},}\ }\href {\doibase
  10.1007/s00220-017-2995-6} {\bibfield  {journal} {\bibinfo  {journal}
  {Commun. Math. Phys.}\ }\textbf {\bibinfo {volume} {356}},\ \bibinfo {pages}
  {1017--1056} (\bibinfo {year} {2017})},\ \Eprint
  {http://arxiv.org/abs/1701.02800} {arXiv:1701.02800 [hep-th]} \BibitemShut
  {NoStop}%
\bibitem [{Note1()}]{Note1}%
  \BibitemOpen
  \bibinfo {note} {In practise we achieve this by using the $F$-symbols of
  \cite {osborne2019fsymbols} with a mild gauge transformation.}\BibitemShut
  {Stop}%
\bibitem [{\citenamefont {{Tetel'man}}(1982)}]{1982JETP...55..306T}%
  \BibitemOpen
  \bibfield  {author} {\bibinfo {author} {\bibfnamefont {M.~G.}\ \bibnamefont
  {{Tetel'man}}},\ }\bibfield  {title} {\enquote {\bibinfo {title} {{Lorentz
  group for two-dimensional integrable lattice systems}},}\ }\href@noop {}
  {\bibfield  {journal} {\bibinfo  {journal} {Soviet Journal of Experimental
  and Theoretical Physics}\ }\textbf {\bibinfo {volume} {55}},\ \bibinfo
  {pages} {306} (\bibinfo {year} {1982})}\BibitemShut {NoStop}%
\bibitem [{\citenamefont {Links}\ \emph {et~al.}(2001)\citenamefont {Links},
  \citenamefont {Zhou}, \citenamefont {McKenzie},\ and\ \citenamefont
  {Gould}}]{links2001ladder}%
  \BibitemOpen
  \bibfield  {author} {\bibinfo {author} {\bibfnamefont {Jon}\ \bibnamefont
  {Links}}, \bibinfo {author} {\bibfnamefont {Huan-Qiang}\ \bibnamefont
  {Zhou}}, \bibinfo {author} {\bibfnamefont {Ross~H}\ \bibnamefont {McKenzie}},
  \ and\ \bibinfo {author} {\bibfnamefont {Mark~D}\ \bibnamefont {Gould}},\
  }\bibfield  {title} {\enquote {\bibinfo {title} {Ladder operator for the
  one-dimensional hubbard model},}\ }\href@noop {} {\bibfield  {journal}
  {\bibinfo  {journal} {Physical review letters}\ }\textbf {\bibinfo {volume}
  {86}},\ \bibinfo {pages} {5096} (\bibinfo {year} {2001})}\BibitemShut
  {NoStop}%
\bibitem [{Note2()}]{Note2}%
  \BibitemOpen
  \bibinfo {note} {For models of range $r$, the auxiliary space contains $r-1$
  copies of the physical space.}\BibitemShut {Stop}%
\bibitem [{Note3()}]{Note3}%
  \BibitemOpen
  \bibinfo {note} {The proof is closely related to the standard `train-track'
  argument of integrability, see e.g. \cite {faddeev1996algebraic}, which
  constructs an RTT relation from the RLL relation and proves $[t(u),t(v)]=0$
  by taking a trace. This has been adapted in \cite {Corcoran:2024ofo} to
  integrable models on constrained Hilbert spaces by inserting appropriate
  projectors.}\BibitemShut {Stop}%
\bibitem [{\citenamefont {Holzhey}\ \emph {et~al.}(1994)\citenamefont
  {Holzhey}, \citenamefont {Larsen},\ and\ \citenamefont
  {Wilczek}}]{Holzhey:1994we}%
  \BibitemOpen
  \bibfield  {author} {\bibinfo {author} {\bibfnamefont {Christoph}\
  \bibnamefont {Holzhey}}, \bibinfo {author} {\bibfnamefont {Finn}\
  \bibnamefont {Larsen}}, \ and\ \bibinfo {author} {\bibfnamefont {Frank}\
  \bibnamefont {Wilczek}},\ }\bibfield  {title} {\enquote {\bibinfo {title}
  {{Geometric and renormalized entropy in conformal field theory}},}\ }\href
  {\doibase 10.1016/0550-3213(94)90402-2} {\bibfield  {journal} {\bibinfo
  {journal} {Nucl. Phys. B}\ }\textbf {\bibinfo {volume} {424}},\ \bibinfo
  {pages} {443--467} (\bibinfo {year} {1994})},\ \Eprint
  {http://arxiv.org/abs/hep-th/9403108} {arXiv:hep-th/9403108} \BibitemShut
  {NoStop}%
\bibitem [{\citenamefont {Calabrese}\ and\ \citenamefont
  {Cardy}(2004)}]{Calabrese:2004eu}%
  \BibitemOpen
  \bibfield  {author} {\bibinfo {author} {\bibfnamefont {Pasquale}\
  \bibnamefont {Calabrese}}\ and\ \bibinfo {author} {\bibfnamefont {John~L.}\
  \bibnamefont {Cardy}},\ }\bibfield  {title} {\enquote {\bibinfo {title}
  {{Entanglement entropy and quantum field theory}},}\ }\href {\doibase
  10.1088/1742-5468/2004/06/P06002} {\bibfield  {journal} {\bibinfo  {journal}
  {J. Stat. Mech.}\ }\textbf {\bibinfo {volume} {0406}},\ \bibinfo {pages}
  {P06002} (\bibinfo {year} {2004})},\ \Eprint
  {http://arxiv.org/abs/hep-th/0405152} {arXiv:hep-th/0405152} \BibitemShut
  {NoStop}%
\bibitem [{Note4()}]{Note4}%
  \BibitemOpen
  \bibinfo {note} {We used a maximum truncation error of $10^{-9}$ and a
  maximum bond dimension of 1700.}\BibitemShut {Stop}%
\bibitem [{\citenamefont {Fishman}\ \emph {et~al.}(2022)\citenamefont
  {Fishman}, \citenamefont {White},\ and\ \citenamefont
  {Stoudenmire}}]{Fishman_2022}%
  \BibitemOpen
  \bibfield  {author} {\bibinfo {author} {\bibfnamefont {Matthew}\ \bibnamefont
  {Fishman}}, \bibinfo {author} {\bibfnamefont {Steven}\ \bibnamefont {White}},
  \ and\ \bibinfo {author} {\bibfnamefont {Edwin}\ \bibnamefont
  {Stoudenmire}},\ }\bibfield  {title} {\enquote {\bibinfo {title} {The itensor
  software library for tensor network calculations},}\ }\href {\doibase
  10.21468/scipostphyscodeb.4} {\bibfield  {journal} {\bibinfo  {journal}
  {SciPost Physics Codebases}\ } (\bibinfo {year} {2022}),\
  10.21468/scipostphyscodeb.4}\BibitemShut {NoStop}%
\bibitem [{\citenamefont {Copetti}\ \emph
  {et~al.}(2024{\natexlab{a}})\citenamefont {Copetti}, \citenamefont
  {Cordova},\ and\ \citenamefont {Komatsu}}]{Copetti:2024rqj}%
  \BibitemOpen
  \bibfield  {author} {\bibinfo {author} {\bibfnamefont {Christian}\
  \bibnamefont {Copetti}}, \bibinfo {author} {\bibfnamefont {Lucia}\
  \bibnamefont {Cordova}}, \ and\ \bibinfo {author} {\bibfnamefont {Shota}\
  \bibnamefont {Komatsu}},\ }\bibfield  {title} {\enquote {\bibinfo {title}
  {{Non-Invertible Symmetries, Anomalies and Scattering Amplitudes}},}\
  }\href@noop {} {\  (\bibinfo {year} {2024}{\natexlab{a}})},\ \Eprint
  {http://arxiv.org/abs/2403.04835} {arXiv:2403.04835 [hep-th]} \BibitemShut
  {NoStop}%
\bibitem [{\citenamefont {Copetti}\ \emph
  {et~al.}(2024{\natexlab{b}})\citenamefont {Copetti}, \citenamefont
  {Cordova},\ and\ \citenamefont {Komatsu}}]{Copetti:2024dcz}%
  \BibitemOpen
  \bibfield  {author} {\bibinfo {author} {\bibfnamefont {Christian}\
  \bibnamefont {Copetti}}, \bibinfo {author} {\bibfnamefont {Lucia}\
  \bibnamefont {Cordova}}, \ and\ \bibinfo {author} {\bibfnamefont {Shota}\
  \bibnamefont {Komatsu}},\ }\bibfield  {title} {\enquote {\bibinfo {title}
  {{S-Matrix Bootstrap and Non-Invertible Symmetries}},}\ }\href@noop {} {\
  (\bibinfo {year} {2024}{\natexlab{b}})},\ \Eprint
  {http://arxiv.org/abs/2408.13132} {arXiv:2408.13132 [hep-th]} \BibitemShut
  {NoStop}%
\bibitem [{\citenamefont {Faddeev}(1996)}]{faddeev1996algebraic}%
  \BibitemOpen
  \bibfield  {author} {\bibinfo {author} {\bibfnamefont {L.}~\bibnamefont
  {Faddeev}},\ }\href@noop {} {\enquote {\bibinfo {title} {How algebraic bethe
  ansatz works for integrable model},}\ } (\bibinfo {year} {1996}),\ \Eprint
  {http://arxiv.org/abs/hep-th/9605187} {arXiv:hep-th/9605187 [hep-th]}
  \BibitemShut {NoStop}%
\end{thebibliography}%

\end{document}